# Correlation between structure and electrical transport in ion-irradiated graphene grown on Cu foils

Grant Buchowicz[1,2], Peter R. Stone[1,2], Jeremy T. Robinson[3], Cory D. Cress[3], Jeffrey W. Beeman[1], Oscar D. Dubon[1,2,(a)]

[1]Lawrence Berkeley National Laboratory, Berkeley, CA 94720;
[2]Department of Materials Science and Engineering, University of California, Berkeley, CA 94720;
[3]Naval Research Laboratory, Washington, D.C. 20375

Abstract:

Graphene grown by chemical vapor deposition and supported on $SiO_2$ and sapphire substrates was studied following controlled introduction of defects induced by 35 keV carbon ion irradiation.  Changes in Raman spectra following fluences ranging from $10^{12}$ $cm^{-2}$ to $10^{15}$ $cm^{-2}$ indicate that the structure of graphene evolves from a highly-ordered layer, to a patchwork of disordered domains, to an essentially amorphous film.  These structural changes result in a dramatic decrease in the Hall mobility by orders of magnitude while, remarkably, the Hall concentration remains almost unchanged, suggesting that the Fermi level is pinned at a hole concentration near $1 \times 10^{13}$ $cm^{-2}$.  A model for scattering by resonant scatterers is in good agreement with mobility measurements up to an ion fluence of $1 \times 10^{14}$ $cm^{-2}$.

(a) email address:  oddubon@berkeley.edu

Graphene, a sheet of $sp^2$-bonded carbon atoms, is attracting tremendous interest due to its potentially transformative impact across a wide range of applications including advanced electronics and sensing [1]. The electronic structure and high quality of isolated, microscale flakes of nearly-perfect graphene enable the observation of spectacular properties such as the quantum Hall effect [2] and carrier mobilities approaching 120,000 cm$^2$/Vs [3] near room temperature. Graphene grown on copper foils by chemical vapor deposition (CVD) has garnered much attention as a method for producing large-area, monolayer-thick films that can be transferred to a variety of supporting substrates [1,4]. Because of the more defective nature of CVD-grown graphene compared to exfoliated flakes, an important challenge remains to understand the role of defects on electronic properties.

Carrier scattering in supported graphene is attributed to a variety of sources including ripples in the graphene layer, point defects and their associated short-range potentials, electron-electron (hole-hole) interactions, charged impurities residing in the supporting substrate, and adsorbed atoms on the surface [5,6,7]. Earlier studies suggested that charged impurity scattering could explain the dominant behavior in experimental findings [5,8]. However, more recent studies indicate that transport is limited by resonant scatterers [9,10], atomic-scale defects such as vacancies or molecules adsorbed to the surface (*e.g.* hydrogen) that generate so-called "midgap states" very close to the Dirac point [11]. By controllably introducing defects into graphene, one may be able to understand how these mechanisms limit transport. Previous studies reported vacancy-type defects induced by ion irradiation of exfoliated graphene [12,13]. In this Letter we report on the irradiation of CVD-grown graphene with carbon ions.  While the carrier mobility depends strongly on ion-induced damage, the carrier density is very insensitive to $C^+$ irradiation even as the graphene becomes highly defective.

Graphene films were grown by CVD onto Cu foils using methane and transferred onto either $SiO_2$/Si or sapphire substrates using the method described by Li *et al.* [14]. The predominant film thickness of one monolayer was identified by atomic force microscopy (AFM) and Raman spectroscopy.  Each sample was irradiated with 35 keV carbon ions at different fluences ranging from $1x10^{12}$ cm$^{-2}$ to $1x10^{15}$ cm$^{-2}$.  Carbon ions of this energy and these fluences were chosen in order to mitigate undesired chemical reactions with the graphene film and to controllably introduce point defects (i.e., avoid cascades that would be associated with heavier ions or higher energies).  The irradiation conditions were chosen such that end-of-range damage would be away from the graphene film.  Based on SRIM simulations, we determined that sputtered atoms from the substrate have negligible effect on the graphene.  In order to estimate the irradiation-induced defect density, we referred to simulations performed by Lehtinen *et al* [15].  Interpreting their results for irradiation by 35 keV $C^+$, we estimate the probability to induce single vacancies, double vacancies or complex resonant scatterers per incoming ion to be within 6-8%.

Raman spectra were collected using a double-resonance Raman process over wavenumbers of 1000 to 3000 cm$^{-1}$ to investigate the structural changes in the material with increasing irradiation.  Variable-temperature resistivity measurements were performed from room temperature down to 5 K while the Hall effect was measured in the same instrument at fields up to 5 Tesla.  Samples were tested in a He gas ambient, which also served as the cooling medium.  Unirradiated graphene films on $SiO_2$ were *p*-type (sheet Hall concentration ~$1\times10^{13}$ cm$^{-2}$) and displayed an average room-temperature Hall mobility of 1200 cm$^2$/Vs.

Raman spectra for graphene-on-$SiO_2$ following select ion fluences are shown in Figure 1a. The unirradiated graphene shows two features: the G line at 1580 cm$^{-1}$ and the more intense G' (or 2D) line appearing at ~2700 cm$^{-1}$.  A third feature, the D line located at 1360 cm$^{-1}$, appears in graphene films that contain lattice disorder [16].  The negligible signal from the D line in the unirradiated sample reflects the relatively low level of disorder in the as-grown films.  After an ion fluence of $10^{12}$ cm$^{-2}$, the D line appears but is very weak compared to the G and G' lines.  Following a fluence of $10^{13}$ cm$^{-2}$, the D line intensity is nearly the same as that of the G line.  Compagnini *et al.* report that when the D line becomes more intense, structurally-disordered domains grow throughout the graphene layer [13].

Following a fluence of $10^{14}$ cm$^{-2}$, the D' line, corresponding to an independent intravalley scattering process, appears at 1620 cm$^{-1}$ [16].  After this fluence, the D line is twice as intense as the G line while the G' line intensity has decreased significantly. After total fluences of 5 x $10^{14}$ cm$^{-2}$ and $10^{15}$ cm$^{-2}$, the G' line is no longer visible, the G and D' lines are indistinguishable from each other, and the D line has broadened due to coalescence of disordered regions [13].  This trend with increasing irradiation is consistent with studies of Lucchese *et al.* performed on exfoliated graphene [17].

The observed evolution of the Raman spectra is consistent with an amorphization process described by Ferrari and Robertson [18] as reflected in Figure 1b.  Region I, where the ratio $I_D$ / $I_G$ is increasing with ion fluence, represents a range of ion-induced damage in which disordered domains grow. As a result, the G' line decreases to zero intensity since increasing disorder prevents second-order processes. In Region II, where the intensity ratio is decreasing, the damage is sufficiently high such that the phonon modes soften and the D and G lines broaden.

Figure 2 shows the temperature dependence of sheet resistance and Hall mobility of graphene-on-SiO$_2$ for ion fluences between 1x10$^{12}$ cm$^{-2}$ and 5x10$^{14}$ cm$^{-2}$. Both the resistance and mobility of the unirradiated film display little to no change over temperature. Irradiation with 10$^{12}$ cm$^{-2}$ produces essentially no change in the sample's metallic behavior but results in a small increase (decrease) in the magnitude of the resistance (mobility). However, after a carbon ion fluence of 5 x 10$^{14}$ cm$^{-2}$, the resistance changes by three orders of magnitude between 300 K and 9 K and scales nearly exponentially with $T^{-1/3}$. This linear relationship between $log(R)$ and $T^{1/3}$ suggests that the highly-irradiated samples display a two-dimensional hopping conductivity previously reported in the behavior of amorphous carbon films [19]. We note that a $T^{1/4}$ dependence cannot be excluded given the limited data. This stronger dependence on temperature with increasing ion dose is reflected in the Hall mobility measurements in Figure 2b.

Figure 3 shows the effect of ion irradiation on the normalized Hall mobility (ratio of sample mobility after irradiation to that prior to irradiation) and the sheet Hall concentration for graphene-on-SiO$_2$ (GSi) and graphene-on-sapphire (GSa) at 290 K. The mobility decreases monotonically with ion fluence and is orders of magnitude smaller at fluences above 1x10$^{14}$ cm$^{-2}$ than the initial mobilities for GSi and GSa. Remarkably, the sheet Hall concentration, irrespective of substrate type, remains largely unchanged upon irradiation and is saturated at a value of ~$1.3\times10^{13}$ cm$^{-2}$, which suggests pinning of the Fermi level.  Similar behavior of the Fermi level in ion-irradiated conventional semiconductors such as GaAs has been observed and is explained by the amphoteric defect model [20] in which the Fermi level is stabilized by native defects.

A simple model for the mobility after irradiation ($\mu_T$) can be constructed using Mattheissen's rule:

$$\frac{1}{\mu_T} = \frac{1}{\mu_0} + \frac{1}{\mu_d} \tag{1}$$

where $\mu_0$ is the mobility measured before irradiation and $\mu_d$ is the mobility contribution from the induced defects.  Assuming that irradiation can create an induced defect concentration of resonant scatterers with a circular potential well of radius R, then the conductivity takes the form

$$\sigma_d = ne\mu_d = \frac{2e^2}{\pi h} \frac{n}{n_d} \left( \ln\left[ R\sqrt{\pi n} \right] \right)^2 \tag{2}$$

where $n$ is the carrier concentration and $n_d$ is the induced defect concentration [21], which we estimate to be about 0.07 times the incoming ion fluence [15].  A model of $\mu_d$ /$\mu_T$ with $R$ = 3.0 Å is plotted in Figure 3a [22].  The model is in good agreement with the experimental data up to an ion fluence of $10^{14}$ cm$^{-2}$.  Above this value, the Hall mobility displays a fluence dependence that is stronger than that expected from the model, indicating that there could be a stronger contribution limiting the mobility in that region as a result of significant disorder or an amorphous-like film structure.

Connecting the electronic characterization with analysis of the Raman data, the damage induced by ion-irradiation in Region I can be well explained by a model identifying resonant scatterers as responsible for the dominant scattering mechanism. While the model shows to be in agreement with the changes in mobility for graphene isolated on both $SiO_2$ and sapphire up to around $10^{14}$ cm$^{-2}$ ion fluence, there must be a different transport mechanism to account for the behavior in Region II. An increasing ion fluence leads to decreasing mobilities with stronger temperature-dependence, uncharacteristic of the contributions from charged impurities or resonant scatterers. Rather, the behavior at extreme levels of disorder in this sample suggests a two-dimensional hopping conduction mechanism like that of an amorphous-like carbon film.

In summary, we have irradiated graphene with carbon ions to induce a defect concentration that reduces mobility significantly with little variability of the sheet carrier concentration. The reduction in mobility can be attributed to resonant scattering for up to an ion fluence near $1 \times 10^{14}$ cm$^{-2}$. The divergence between the model and the experimental data at higher fluences marks the transition from structurally disordered graphene to an amorphous-like carbon film.

The work at the Lawrence Berkeley National Laboratory (ion irradiation and electrical characterization of graphene on $SiO_2$) was supported by the Director, Office of Science, Office of Basic Energy Sciences, and Division of Materials Sciences and Engineering of the U.S. Department of Energy under Contract No. De-Ac02-05Ch11231. O. D. D. acknowledges support from the National Science Foundation under contract number DMR-0349257 for electrical measurements of graphene on sapphire. This work was supported in part by the Office of Naval Research, NRL's Nanoscience Institute, and the Defense Threat Reduction Agency under MIPR No. 10-2197M.

Figure Captions:

FIG. 1. (a) Raman spectra for graphene-on-SiO$_2$ at select ion fluences. (b) The ratio of the intensities for the D and G lines is plotted against the ion fluence for the spectra shown in Fig. 1a.

FIG. 2. (a) Sheet resistance for graphene-on-SiO2 versus $T^{-1/3}$ for select carbon ion fluences. At higher irradiation, resistance is fitted to a relation of the form $R = R_0$ exp $[(T_0/T)^{1/3}]$. (b) Hall mobility for graphene-on-SiO$_2$ versus temperature for select carbon ion fluences.

FIG. 3. (a) Normalized Hall mobility and (b) sheet Hall concentration of graphene-on-SiO$_2$ (GSi) and graphene-on-sapphire (GSa) versus the C ion fluence. A model for mobility based on a resonant scatterer (RS) defect contribution with R = 3.0 Å is also shown in (a). Initial mobilities for GSi and GSa are ~1200 cm$^2$/Vs and ~500 cm$^2$/Vs, respectively.

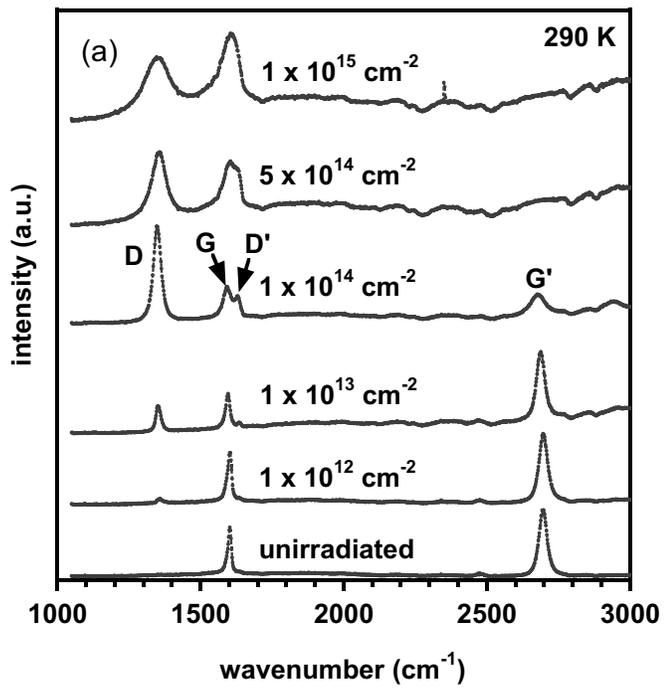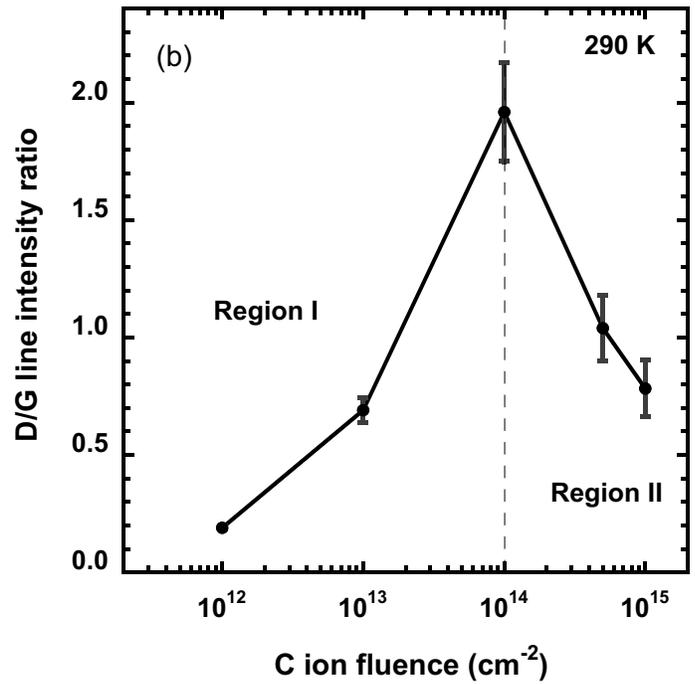

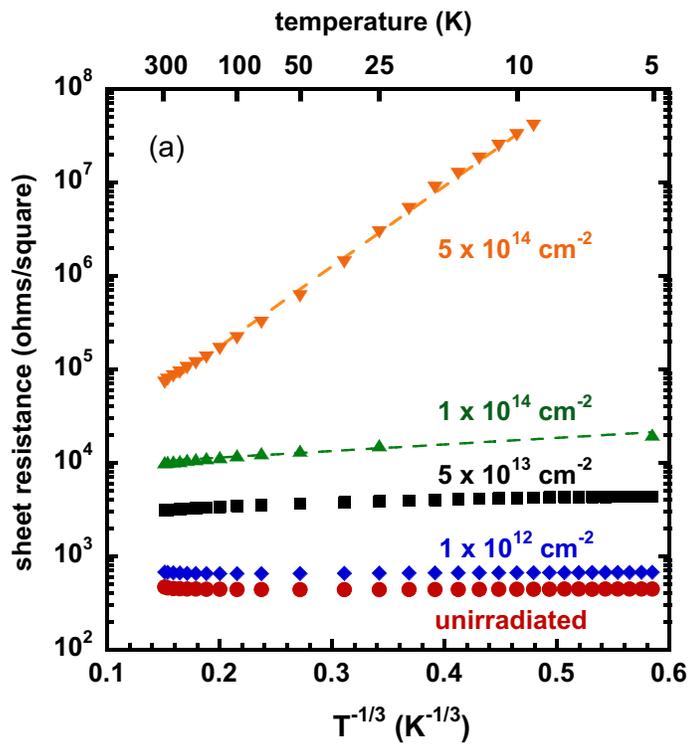
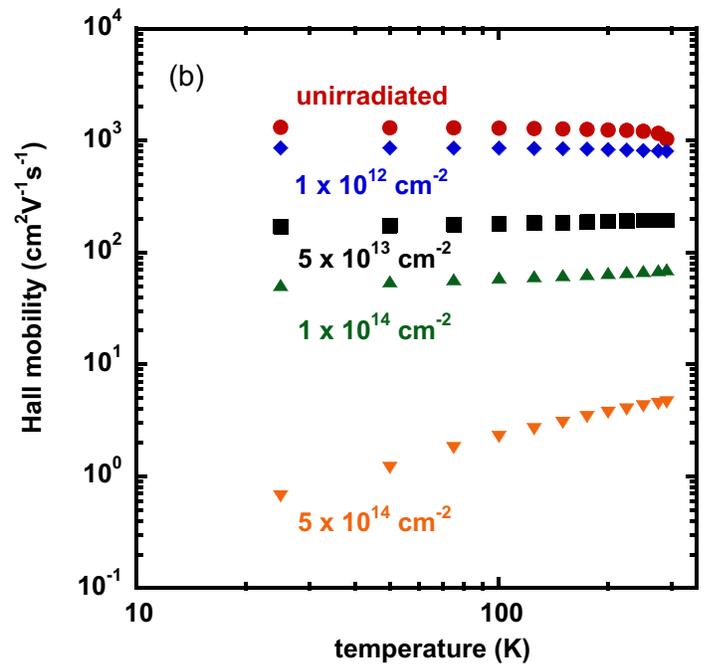

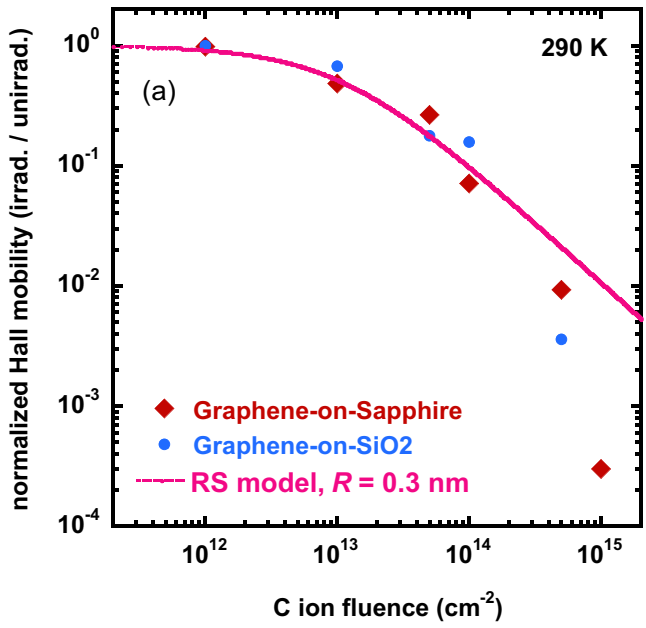

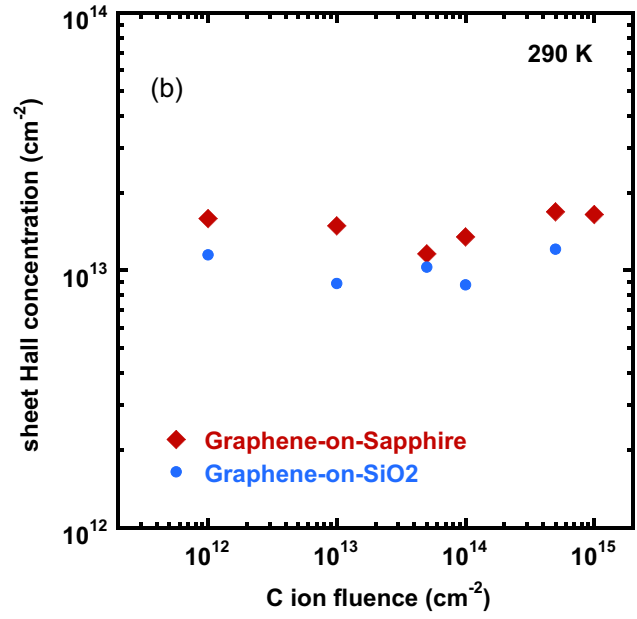